\documentclass[11pt]{article}           
\usepackage{amsmath}

\newtheorem{Theorem}{Theorem}
\newtheorem{Remark}{Remark}

\usepackage{graphics}

\date{}
\title{A Novel Symmetry Constraint Of The Super cKdV System}
\author{JING YU$^{a,b}$, JINGSONG HE$^{c,b}$\footnote{Corresponding author, E-mail
address:hejingsong@nbu.edu.cn;jshe@ustc.edu.cn}, YI CHENG$^{b}$,
JINGWEI HAN$^{a}$
\vspace{4mm}\\
$^{a}$School of Science, Hangzhou Dianzi University, Hangzhou,
Zhejiang, 310018,  China\\
$^{b}$Department of Mathematics, University of Science and
Technology of China, Hefei, Anhui,\\ 230026,  China\\
$^{c}$Department of Mathematics, Ningbo University, Ningbo,
Zhejiang, 315211,
 China}

\begin{document}
 \maketitle

 \begin{abstract}

 A new (1+1)-dimensional integrable system, i. e. the super coupled Korteweg-de Vries (cKdV) system,
has been  constructed by a super extension of the well-known (1+1)-dimensional cKdV system. For this new system,
a novel symmetry constraint between the potential and eigenfunction can be obtained by means of the binary
nonlinearization of its Lax pairs. The constraints for even variables are explicit and the constraints for
odd variables are implicit.  Under the symmetry constraint, the spacial part and the temporal
 parts of the equations associated with the Lax pairs for the super cKdV system
can be decomposed into the super finite-dimensional integrable
Hamiltonian systems on the supersymmetry manifold $R^{4N|2N+2}$,
whose integrals of motion are explicitly given.


\noindent{\bf PACS codes(2008):}\ 02.30.Ik, 02.90.+p
\end{abstract}

\section{Introduction}

The super-extensions of the classical integrable systems  lead to
super integrable systems and they have undergone extensive
development in the past years. There are many super integrable
systems in literatures, such as the super AKNS system
\cite{Gurses1}-\cite{Lizhang2}, the super KdV equation
\cite{Kupershmidt}-\cite{Shaw1}, the super KP hierarchy
\cite{Manin}-\cite{Shaw2}, etc. It was known that super systems
contained the odd variables which would provide more prolific fields
for mathematical researchers and physical ones. Darboux
transformation \cite{Liu1}-\cite{Siddiq}, bi-Hamiltonian structure
\cite{Popowicz}-\cite{Kersten}, Painle$\acute{v}$e analysis
\cite{Mathieu} and so on, have been widely studied. Very recently,
nonlinearization of the super AKNS system and the super Dirac system
have been investigated in Refs.
\cite{HeYuZhouCheng}-\cite{YuHeMaCheng}.

It is well known that mono-nonlinearization technique was firstly
proposed by Cao in Ref. \cite{Cao1}, and binary-nonlinearization
technique was proposed by Ma in Ref. \cite{Ma1}. Both
mono-nonlinearization and binary-nonlinearization have the following
characteristics. Firstly, the advantage of nonlinearizaton method is
to decompose infinite dimensional systems into finite ones.
Secondly, one of the essential steps of nonlinearization method is
to calculate the variational derivative. Lastly, the key of the
nonlinearization method is to find symmetry constraints between the
potential and the eigenfunction by means of variational derivative.
On the one hand, nonlinearization of Lax pairs is valid for many
classical integrable systems \cite{Lizeng}-\cite{Zhouruguang}. On
the other hand, binary nonlinearization has been applied to the
super AKNS system and the super Dirac system in
Refs.\cite{HeYuZhouCheng}-\cite{YuHeMaCheng}. However, is
nonlinearization method valid for the other super integrable
systems? For the cKdV system, the answer is affirmative in this
paper. The cKdV system firstly proposed by Hirota and Satsuma in
Ref.\cite{HirotaSatsuma} is very important in the classical
integrable systems. Its mono-nonlinearization and Darboux
transformation were studied in Refs.\cite{Cao1990, Qin2004}.

The paper is organized as follows. In the next section, the cKdV
system  is to be extended into the super one, and the super
Hamiltonian structure will be obtained for new system by means of
the supertrace identity. In section 3, variational derivative of the
spectral parameter with respect to the potential is calculated by
Lemma 2.1 in Ref. \cite{YuHeMaCheng}, and a symmetry constraint
between the potential and the eigenfunction can be found. The
symmetry constraint is an interesting constraint, and it is explicit
for even elements, but it is implicit for odd elements. Then in
section 4, after introduction of two new odd variables, the novel
symmetry constraint is substituted into the Lax pairs and the
adjoint Lax pairs of the super cKdV system while considering the two
new  variables. And we find that the constrained Lax pairs and the
adjoint Lax pairs of the super cKdV system are super Hamiltonian
systems, and are completely integrable systems in the Liouville
sense. Integrals of motion with odd eigenfunctions are given
explicitly. The conclusions and discussions are given in section 5.

\small \baselineskip 13pt

\section{The super cKdV soliton hierarchy}

Let's begin with the following spectral problem
\begin{equation}\label{c1}
\phi_x=U(u, \lambda)\phi,\quad U(u,
\lambda)=\left(\begin{array}{ccc}
-\frac{1}{2}\lambda+\frac{1}{2}q&-r&\alpha\\
1&\frac{1}{2}\lambda-\frac{1}{2}q&\beta\\
\beta&-\alpha&0\end{array}\right),\quad
u=\left(\begin{array}{c}q\\r\\\alpha\\\beta\end{array}\right),\quad
\phi=\left(\begin{array}{c}\phi_1\\\phi_2\\\phi_3\end{array}\right),
\end{equation}
where $u$ is a potential, and $\lambda$ is a spectral parameter. Set
$p(q)=p(r)=p(\lambda)=0$, and $p(\alpha)=p(\beta)=1$. Here $p(f)$
means the parity of arbitrary function $f$. Note that $U \in
\mathbf{B}(0,1)$, where $\mathbf{B}(0,1)$ is a  Lie superalegra.

Set
$$V=\left(\begin{array}{ccc}
A&B&\rho\\C&-A&\delta\\\delta&-\rho&0\end{array}\right)
$$
where $p(A)=p(B)=p(C)=0$, $p(\rho)=p(\delta)=1$. Noting that
$$UV-VU=\left(\begin{array}{ccc}
-B-rC+\alpha\delta+\beta\rho&-\lambda
B+2rA+qB-2\alpha\rho&-\frac{1}{2}\lambda\rho-\alpha A-\beta
B+\frac{1}{2}q\rho-r\delta\\
\lambda
C+2A-qC+2\beta\delta&B+rC-\alpha\delta-\beta\rho&\frac{1}{2}\delta+\beta
A-\alpha C+\rho-\frac{1}{2}q\delta\\\frac{1}{2}\delta+\beta A-\alpha
C+\rho-\frac{1}{2}q\delta&\frac{1}{2}\lambda\rho+\alpha A+\beta
B-\frac{1}{2}q\rho+r\delta&0 \end{array}\right),$$
then there goes co-adjoint representation equation
\begin{equation}
V_x=[U, V]=UV-VU,
\end{equation}
it becomes
\begin{equation}\label{c2}\left\{\begin{array}{l}
A_x=-B-rC+\alpha\delta+\beta\rho,\\
B_x=-\lambda B+2rA+qB-2\alpha\rho,\\
C_x=\lambda C+2A-qC+2\beta\delta,\\
\rho_x=-\frac{1}{2}\lambda\rho-\alpha A-\beta
B+\frac{1}{2}q\rho-r\delta,\\
\delta_x=\frac{1}{2}\delta+\beta A-\alpha C+\rho-\frac{1}{2}q\delta.
\end{array}\right.\end{equation}
On setting $A=\sum\limits_{j\geq0}A_j\lambda^{-j}$,
$B=\sum\limits_{j\geq0}B_j\lambda^{-j}$,
$C=\sum\limits_{j\geq0}C_j\lambda^{-j}$,
$\rho=\sum\limits_{j\geq0}\rho_j\lambda^{-j}$,
$\delta=\sum\limits_{j\geq0}\delta_j\lambda^{-j}$, then
equation (\ref{c2}) is equivalent to
\begin{equation}\label{c3}\left\{\begin{array}{l}
B_0=C_0=\rho_0=\delta_0=0,\\
A_{j, x}=-B_j-rC_j+\beta\rho_j+\alpha\delta_j,\quad j\geq0,\\
B_{j, x}=-B_{j+1}+2rA_j+qB_j-2\alpha\rho_j,\quad j\geq0,\\
C_{j, x}=C_{j+1}+2A_j-qC_j+2\beta\delta_j,\quad j\geq0,\\
\rho_{j, x}=-\frac{1}{2}\rho_{j+1}-\alpha A_j-\beta
B_j+\frac{1}{2}q\rho_j-r\delta_j,\quad j\geq0,\\
\delta_{j, x}=\frac{1}{2}\delta_{j+1}+\beta A_j-\alpha
C_j+\rho_j-\frac{1}{2}q\delta_j,\quad j\geq0.\end{array}\right.
\end{equation}
It can be written as the following recurrence relation
\begin{equation}\label{c6}\left(\begin{array}{c}
A_{n+1}\\-C_{n+1}\\2\delta_{n+1}\\-2\rho_{n+1}\end{array}\right)
={\cal
L}\left(\begin{array}{c}A_n\\-C_n\\2\delta_n\\-2\rho_n\end{array}\right),\end{equation}
where the recursive operator is given by
$${\cal L}=\left(\begin{array}{cccc}
-\partial+\partial^{-1}q\partial&r+\partial^{-1}r\partial&
\frac{1}{2}\alpha+\partial^{-1}\alpha\partial&-\frac{1}{2}\beta+\partial^{-1}\beta\partial\\
2&\partial+q&\beta&0\\
-4\beta&-4\alpha&2\partial+q&2\\
-4\beta\partial+4\alpha&4r\beta&2r-2\alpha\beta&-2\partial+q\end{array}\right),$$
with $\partial=d/dx$ and
$\partial\partial^{-1}=\partial^{-1}\partial=1$.

Owing to $B_0=C_0=\rho_0=\delta_0=0$, we get that $A_{0, x}=0$. So
we choose the initial value $A_0=-\frac{1}{2}$. If we set all
constants of integration to be zero, all $A_j, B_j, C_j, \rho_j,
\delta_j (j>0)$ are uniquely given by (\ref{c6}). For instance
$$A_1=0, B_1=-r,  C_1=1, \rho_1=\alpha, \delta_1=\beta,$$
$$A_2=-r+2\alpha\beta, B_2=r_x-q r, C_2=q, \rho_2=-2\alpha_x+q\alpha,
\delta_2=2\beta_x+q\beta.$$

Then, consider the auxiliary spectral problem associated with the spectral
problem (\ref{c1})
\begin{equation}\label{c4}
\phi_{t_n}=V^{(n)}\phi,\end{equation} where
$$V^{(n)}=(\lambda^{n}V)_++\Delta_n=\sum_{j=0}^{n}\left(\begin{array}{ccc}
A_j&B_j&\rho_j\\C_j&-A_j&\delta_j\\\delta_j&-\rho_j&0\end{array}\right)\lambda^{n-j}
+\left(\begin{array}{ccc}
\frac{1}{2}C_{n+1}&0&0\\0&-\frac{1}{2}C_{n+1}&0\\0&0&0\end{array}\right),$$
 and $(\lambda^{n}V)_+$  denotes non-negative power of $\lambda$ in $V$.

The compatibility conditions of Lax pairs
\begin{equation}\label{c27}
\phi_x=U\phi,\quad \phi_{t_n}=V^{(n)}\phi,
\end{equation}
determine a hierarchy of super cKdV system
\begin{equation}\label{c5}\left\{\begin{array}{l}
q_{t_n}=C_{n+1, x},\\
r_{t_n}=B_{n+1}+rC_{n+1},\\
\alpha_{t_n}=\frac{1}{2}\alpha C_{n+1}-\frac{1}{2}\rho_{n+1},\\
\beta_{t_n}=\frac{1}{2}\delta_{n+1}-\frac{1}{2}\beta
C_{n+1}.\end{array}\right.\end{equation} The first nonlinear cKdV
system in the hierarchy (\ref{c5}) reads as
\begin{equation}\left\{\begin{array}{l}
q_{t_2}=q_{xx}+2qq_x+2r_x-4\alpha_x\beta-4\alpha\beta_x-4\beta\beta_{xx},\\
r_{t_2}=-r_{xx}+2q_xr+2qr_x+4\alpha\alpha_x-4r\beta\beta_x,\\
\alpha_{t_2}=-2\alpha_{xx}+\frac{3}{2}q_x\alpha+2q\alpha_x+r_x\beta+2r\beta_x
-2\alpha\beta\beta_x,\\
\beta_{t_2}=2\beta_{xx}+\frac{1}{2}q_x\beta+2q\beta_x+2\alpha_x,
\end{array}\right.\end{equation}
whose Lax pairs are $U$ and
$$V^{(2)}=\left(\begin{array}{ccc}
-\frac{1}{2}\lambda^{2}+\frac{1}{2}q_x+\frac{1}{2}q^{2}-2\beta\beta_x&
-r\lambda+r_x-qr&\alpha\lambda-2\alpha_x+q\alpha\\ \lambda+q&
\frac{1}{2}\lambda^{2}-\frac{1}{2}q_x-\frac{1}{2}q^{2}+2\beta\beta_x&
\beta\lambda+2\beta_x+q\beta\\ \beta\lambda+2\beta_x+q\beta&
-\alpha\lambda+2\alpha_x-q\alpha&0 \end{array}\right).$$

In what follows,  the super Hamiltonian structures of the super cKdV system (\ref{c5})
can be achieved. Using  the
super trace identity \cite{Hu, Maqin}
\begin{equation}\label{c28}
\frac{\delta}{\delta u}\int Str(V\frac{\partial
U}{\partial\lambda})dx=(\lambda^{-\gamma}\frac{\partial}{\partial\lambda}\lambda^{\gamma})
Str(\frac{\partial U}{\partial u}V),\end{equation} where Str means
super trace,  we have
$$\left(\begin{array}{c}
\frac{\delta}{\delta q}\\\frac{\delta}{\delta
r}\\\frac{\delta}{\delta \alpha}\\\frac{\delta}{\delta
\beta}\end{array}\right)\int-A_{n+1}dx=(\gamma-n)
\left(\begin{array}{c}A_n\\-C_n\\2\delta_n\\-2\rho_n\end{array}\right),$$
where $\gamma$ is an arbitrary constant. Let $n=1$ in above
equality, we obtain $\gamma=0$. Therefore, we get the following
identity
$$\left(\begin{array}{c}A_{n+1}\\-C_{n+1}\\2\delta_{n+1}\\-2\rho_{n+1}\end{array}\right)
=\frac{\delta}{\delta u}H_n,\quad H_n=\int\frac{1}{n+1}A_{n+2}dx.$$

Thus, the super cKdV hierarchy can be written as the following super
Hamiltonian form
\begin{equation}\label{c7}
u_{t_n}=\left(\begin{array}{c}q\\r\\\alpha\\\beta\end{array}\right)_{t_n}
=K_n=J\left(\begin{array}{c}A_{n+1}\\-C_{n+1}\\2\delta_{n+1}\\-2\rho_{n+1}\end{array}\right)
=J\frac{\delta H_n}{\delta u},\end{equation} where the super
symplectic operator is given by
$$J=\left(\begin{array}{cccc}0&-\partial&0&0\\
-\partial&0&\frac{1}{2}\alpha&-\frac{1}{2}\beta\\
0&-\frac{1}{2}\alpha&0&\frac{1}{4}\\
0&\frac{1}{2}\beta&\frac{1}{4}&0\end{array}\right).$$

\section{A novel symmetry constraint}

In this section,  a symmetry constraint between the potential and the eigenfunction can be obtained.
To this end, consider the
adjoint spectral problem associated with spectral problem (\ref{c1})
\begin{equation}\label{c8}
\psi_x=-(U(u, \lambda))^{St}\psi=\left(\begin{array}{ccc}
\frac{1}{2}\lambda-\frac{1}{2}q&-1&\beta\\
r&-\frac{1}{2}\lambda+\frac{1}{2}q&-\alpha\\
-\alpha&-\beta&0\end{array}\right)\psi,\quad
\psi=\left(\begin{array}{c}\psi_1\\\psi_2\\\psi_3\end{array}\right),
\end{equation}
where $St$ means super-transposition.

Using Lemma 2.1 in \cite{YuHeMaCheng}, we can easily get the
variational derivative of the spectral parameter $\lambda$ with
respect to the potential $u$:
\begin{equation}\label{c9}
\frac{\delta\lambda}{\delta u}=\frac{1}{E}\left(\begin{array}{c}
\frac{1}{2}(\psi_1\phi_1-\psi_2\phi_2)\\-\psi_1\phi_2\\\psi_1\phi_3+\psi_3\phi_2\\
\psi_2\phi_3-\psi_3\phi_1\end{array}\right),\end{equation} where
$E=\int\frac{1}{2}(\psi_1\phi_1-\psi_2\phi_2)dx$. When zero boundary
conditions
$\lim_{|x|\rightarrow\infty}\phi=\lim_{|x|\rightarrow\infty}\psi=0$
are imposed,  it satisfies following  equation
\begin{equation}\label{c10}
{\cal L}\frac{\delta\lambda}{\delta
u}=\lambda\frac{\delta\lambda}{\delta u},\end{equation} where ${\cal
L}$ is defined as in (\ref{c6}). The above variational derivative
will serve as a conserved covariant yielding a specific symmetry
used in symmetry constraints.

For Lax pairs (\ref{c27}), we choose the following symmetry
constraint
\begin{equation}\label{c10`}\left(\begin{array}{c}
-r+2\alpha\beta\\-q\\4\beta_x+2q\beta\\4\alpha_x-2q\alpha\end{array}\right)
=\left(\begin{array}{c} \frac{1}{2}(<\Psi_1, \Phi_1>-<\Psi_2,
\Phi_2>)\\-<\Psi_1, \Phi_2>\\<\Psi_1, \Phi_3>+<\Psi_3, \Phi_2>\\
<\Psi_2, \Phi_3>-<\Psi_3, \Phi_1>\end{array}\right),\end{equation}
where $\Phi_i=(\phi_{i1}, \cdots, \phi_{iN})^{T}$,
$\Psi_i=(\psi_{i1}, \cdots, \psi_{iN})^{T} (i=1, 2, 3)$, and $<.,.>$
denotes the standard inner product in $R^{N}$. We find that the odd
potentials $\alpha$ and $\beta$ can not be expressed by
eigenfunctions explicitly, but the even potentials $q$ and $r$ can
be expressed by eigenfunctions explicitly. Therefore, the symmetry
constraint (\ref{c10`}) is a novel constraint.

\begin{Remark}
In classical integrable systems, symmetry constraint between
potential and eigenfunction is either explicit or implicit. To this
day, we haven't got  an example with its  symmetry constraint that could
combine explicit constraint and implicit constraint. Even in super
integrable systems, we haven't got it too. Therefore, eq.(\ref{c10`}) is absolutely a  novel
symmetry constraint.
\end{Remark}

Then denote  the expression of $P(u)$ under the symmetry constraint
(\ref{c10`}) by $\tilde{P}$. From the property (\ref{c10}) and the
recurrence relation (\ref{c6}), we obtain
\begin{equation}\label{c13}\left\{\begin{array}{l}
\tilde{A}_{n+1}=\frac{1}{2}(<\Lambda^{n-1}\Psi_1,
\Phi_1>-<\Lambda^{n-1}\Psi_2, \Phi_2>),\quad n\geq1,\\
\tilde{B}_{n+1}=<\Lambda^{n-1}\Psi_2, \Phi_1>,\quad n\geq1,\\
\tilde{C}_{n+1}=<\Lambda^{n-1}\Psi_1, \Phi_2>,\quad n\geq1,\\
\tilde{\rho}_{n+1}=-\frac{1}{2}(<\Lambda^{n-1}\Psi_2,
\Phi_3>-<\Lambda^{n-1}\Psi_3, \Phi_1>),\quad n\geq1,\\
\tilde{\delta}_{n+1}=\frac{1}{2}(<\Lambda^{n-1}\Psi_1,
\Phi_3>+<\Lambda^{n-1}\Psi_3, \Phi_2>),\quad
n\geq1,\end{array}\right.\end{equation} where
$\Lambda=diag(\lambda_1, \lambda_2, \cdots, \lambda_N).$

\section{Binary nonlinearization}

In the last section, we have found a novel symmetry constraint
(\ref{c10`}). Because the odd potentials $\alpha$ and $\beta$ can
not be explicitly expressed by eigenfunctions, we introduce the
following new independent odd variables
\begin{equation}\label{c23}
\phi_{N+1}=\alpha,\quad\psi_{N+1}=4\beta.
\end{equation}
Choosing N distinct parameters $\lambda_1, \cdots, \lambda_N$, we
obtain the following two spatial and temporal systems
\begin{equation}\label{cspatial}\left\{\begin{array}{l}\left(\begin{array}{c}
\phi_{1j}\\\phi_{2j}\\\phi_{3j}\end{array}\right)_x=U(u, \lambda_j)
\left(\begin{array}{c}
\phi_{1j}\\\phi_{2j}\\\phi_{3j}\end{array}\right),\quad j=1, 2,
\cdots, N,\\
\left(\begin{array}{c}
\psi_{1j}\\\psi_{2j}\\\psi_{3j}\end{array}\right)_x=-U^{St}(u,
\lambda_j) \left(\begin{array}{c}
\psi_{1j}\\\psi_{2j}\\\psi_{3j}\end{array}\right),\quad j=1, 2,
\cdots, N,\end{array}\right.\end{equation}
\begin{equation}\label{ctime}\left\{\begin{array}{l}\left(\begin{array}{c}
\phi_{1j}\\\phi_{2j}\\\phi_{3j}\end{array}\right)_{t_n}=V^{(n)}(u,
\lambda_j) \left(\begin{array}{c}
\phi_{1j}\\\phi_{2j}\\\phi_{3j}\end{array}\right),\quad j=1, 2,
\cdots, N,\\
\left(\begin{array}{c}
\psi_{1j}\\\psi_{2j}\\\psi_{3j}\end{array}\right)_{t_n}=-(V^{(n)})^{St}(u,
\lambda_j) \left(\begin{array}{c}
\psi_{1j}\\\psi_{2j}\\\psi_{3j}\end{array}\right),\quad j=1, 2,
\cdots, N.\end{array}\right.\end{equation} It is easy to verify that
the compatibility condition of (\ref{cspatial}) and (\ref{ctime}) is
still the $n$th super cKdV systems $u_{t_n}=K_n$. When the symmetry
constraint (\ref{c10`}) and new independent variables (\ref{c23})
are considered, systems (\ref{cspatial}) and (\ref{ctime}) become
the following finite-dimensional system
\begin{equation}\label{c11}\left\{\begin{array}{l}
\phi_{1j, x}=\frac{1}{2}(-\lambda_j+<\Psi_1,
\Phi_2>)\phi_{1j}+\frac{1}{2}(<\Psi_1, \Phi_1>-<\Psi_2,
\Phi_2>-\phi_{N+1}\psi_{N+1})\phi_{2j}\\\qquad\quad+\phi_{N+1}\phi_{3j},\\
\phi_{2j, x}=\phi_{1j}+\frac{1}{2}(\lambda_j-<\Psi_1,
\Phi_2>)\phi_{2j}+\frac{1}{4}\psi_{N+1}\phi_{3j},\\
\phi_{3j, x}=\frac{1}{4}\psi_{N+1}\phi_{1j}-\phi_{N+1}\phi_{2j},\\
\phi_{N+1, x}=\frac{1}{4}(<\Psi_2, \Phi_3>-<\Psi_3,
\Phi_1>)+\frac{1}{2}<\Psi_1, \Phi_2>\phi_{N+1},\\
\psi_{1j, x}=\frac{1}{2}(\lambda_j-<\Psi_1,
\Phi_2>)\psi_{1j}-\psi_{2j}+\frac{1}{4}\psi_{N+1}\psi_{3j},\\
\psi_{2j, x}=\frac{1}{2}(-<\Psi_1, \Phi_1>+<\Psi_2,
\Psi_2>+\phi_{N+1}\psi_{N+1})\psi_{1j}+\frac{1}{2}(-\lambda_j+<\Psi_1,
\Phi_2>)\psi_{2j}\\\qquad\quad-\phi_{N+1}\psi_{3j},\\
\psi_{3j, x}=-\phi_{N+1}\psi_{1j}-\frac{1}{4}\psi_{N+1}\psi_{2j},\\
\psi_{N+1, x}=<\Psi_1, \Phi_3>+<\Psi_3, \Phi_2>-\frac{1}{2}<\Psi_1,
\Phi_2>\psi_{N+1},\end{array}\right.\end{equation} where  $1\leq
j\leq N$. Then  system (\ref{c11}) can be written as follows
\begin{equation}\label{c12}\left\{\begin{array}{l}
\Phi_{1, x}=\frac{1}{2}(-\Lambda+<\Psi_1,
\Phi_2>)\Phi_1+\frac{1}{2}(<\Psi_1, \Phi_1>-<\Psi_2,
\Phi_2>-\phi_{N+1}\psi_{N+1})\Phi_2\\\qquad\quad+\phi_{N+1}\Phi_3=\frac{\partial
H_1}{\partial\Psi_1},\\
\Phi_{2, x}=\Phi_1+\frac{1}{2}(\Lambda-<\Psi_1,
\Phi_2>)\Phi_2+\frac{1}{4}\psi_{N+1}\Phi_3=\frac{\partial
H_1}{\partial\Psi_2},\\
\Phi_{3,
x}=\frac{1}{4}\psi_{N+1}\Phi_1-\phi_{N+1}\Phi_2=\frac{\partial
H_1}{\partial\Psi_3},\\
\phi_{N+1, x}=\frac{1}{4}(<\Psi_2, \Phi_3>-<\Psi_3,
\Phi_1>)+\frac{1}{2}<\Psi_1, \Phi_2>\phi_{N+1}=\frac{\partial
H_1}{\partial\psi_{N+1}},\\
\Psi_{1, x}=\frac{1}{2}(\Lambda-<\Psi_1,
\Phi_2>)\Psi_1-\Psi_2+\frac{1}{4}\psi_{N+1}\Psi_3=-\frac{\partial
H_1}{\partial\Phi_1},\\
\Psi_{2, x}=\frac{1}{2}(-<\Psi_1, \Phi_1>+<\Psi_2,
\Phi_2>+\phi_{N+1}\psi_{N+1})\Psi_1+\frac{1}{2}(-\Lambda+<\Psi_1,
\Phi_2>)\Psi_2\\\qquad\quad-\phi_{N+1}\Psi_3=-\frac{\partial
H_1}{\partial\Phi_2},\\
\Psi_{3,
x}=-\phi_{N+1}\Psi_1-\frac{1}{4}\psi_{N+1}\Psi_2=\frac{\partial
H_1}{\partial\Phi_3},\\
\psi_{N+1, x}=<\Psi_1, \Phi_3>+<\Psi_3, \Phi_2>-\frac{1}{2}<\Psi_1,
\Phi_2>\psi_{N+1}=\frac{\partial
H_1}{\partial\phi_{N+1}},\end{array}\right.\end{equation} where
Hamiltonian function
\begin{eqnarray*}
H_1&=&-\frac{1}{2}<\Lambda\Psi_1, \Phi_1>+\frac{1}{2}<\Lambda\Psi_2,
\Phi_2>+\frac{1}{2} <\Psi_1, \Phi_2>(<\Psi_1, \Phi_1>-<\Psi_2,
\Phi_2>)\\&&+<\Psi_2,
\Phi_1>-\frac{1}{2}\phi_{N+1}\psi_{N+1}<\Psi_1,
\Phi_2>+\phi_{N+1}(<\Psi_1, \Phi_3>+<\Psi_3,
\Phi_2>)\\&&+\frac{1}{4}\psi_{N+1}(<\Psi_2, \Phi_3>-<\Psi_3,
\Phi_1>).\end{eqnarray*}

For $t_2$-part, we have the following spectral problem
\begin{equation}\label{t2-1}
\phi_{t_2}=V^{(2)}\phi=\left(\begin{array}{ccc}
-\frac{1}{2}\lambda^{2}+\frac{1}{2}q_x+\frac{1}{2}q^{2}-2\beta\beta_x&
-r\lambda+r_x-qr&\alpha\lambda-2\alpha_x+q\alpha\\ \lambda+q&
\frac{1}{2}\lambda^{2}-\frac{1}{2}q_x-\frac{1}{2}q^{2}+2\beta\beta_x&
\beta\lambda+2\beta_x+q\beta\\ \beta\lambda+2\beta_x+q\beta&
-\alpha\lambda+2\alpha_x-q\alpha&0
\end{array}\right)\phi,\end{equation}
and its adjoint spectral problem
\begin{equation}\label{t2-2}
\psi_{t_2}=-(V^{(2)})^{St}\psi=\left(\begin{array}{ccc}
\frac{1}{2}\lambda^{2}-\frac{1}{2}q_x-\frac{1}{2}q^{2}+2\beta\beta_x&
-\lambda-q&\beta\lambda+2\beta_x+q\beta\\
r\lambda-r_x+qr&-\frac{1}{2}\lambda^{2}+\frac{1}{2}q_x+\frac{1}{2}q^{2}-2\beta\beta_x&
-\alpha\lambda+2\alpha_x-q\alpha\\
-\alpha\lambda+2\alpha_x-q\alpha& -\beta\lambda-2\beta_x-q\beta&0
\end{array}\right)\psi.\end{equation}
Considering N copies of (\ref{t2-1}) and (\ref{t2-2}) under the
symmetry constraint (\ref{c10`}), we obtain the following
finite-dimensional system
\begin{equation}\label{t2-3}\left\{\begin{array}{l}
\phi_{1j,
t_2}=(-\frac{1}{2}\lambda_j^{2}+\frac{1}{2}\tilde{q}_x+\frac{1}{2}\tilde{q}^{2}
-2\tilde{\beta}\tilde{\beta}_x)\phi_{1j}+
(-\tilde{r}\lambda_j+\tilde{r}_x-\tilde{q}\tilde{r})\phi_{2j}
+(\tilde{\alpha}\lambda_j-2\tilde{\alpha}_x+\tilde{q}\tilde{\alpha})\phi_{3j},\\
\phi_{2j, t_2}=(\lambda_j+\tilde{q})\phi_{1j}+
(\frac{1}{2}\lambda_j^{2}-\frac{1}{2}\tilde{q}_x-\frac{1}{2}\tilde{q}^{2}
+2\tilde{\beta}\tilde{\beta}_x)\phi_{2j}
+(\tilde{\beta}\lambda_j+2\tilde{\beta}_x+\tilde{q}\tilde{\beta})\phi_{3j},\\
\phi_{3j,
t_2}=(\tilde{\beta}\lambda_j+2\tilde{\beta}_x+\tilde{q}\tilde{\beta})\phi_{1j}
+(-\tilde{\alpha}\lambda_j+2\tilde{\alpha}_x-\tilde{q}\tilde{\alpha})\phi_{2j},\\
\psi_{1j,
t_2}=(\frac{1}{2}\lambda_j^{2}-\frac{1}{2}\tilde{q}_x-\frac{1}{2}\tilde{q}^{2}
+2\tilde{\beta}\tilde{\beta}_x)\psi_{1j}-(\lambda_j+\tilde{q})\psi_{2j}+
(\tilde{\beta}\lambda_j+2\tilde{\beta}_x+\tilde{q}\tilde{\beta})\psi_{3j},\\
\psi_{2j,
t_2}=(\tilde{r}\lambda_j-\tilde{r}_x+\tilde{q}\tilde{r})\psi_{1j}
+(-\frac{1}{2}\lambda_j^{2}+\frac{1}{2}\tilde{q}_x+\frac{1}{2}\tilde{q}^{2}
-2\tilde{\beta}\tilde{\beta}_x)\psi_{2j}
+(-\tilde{\alpha}\lambda_j+2\tilde{\alpha}_x-\tilde{q}\tilde{\alpha})\psi_{3j},\\
\psi_{3j,
t_2}=(-\tilde{\alpha}\lambda_j+2\tilde{\alpha}_x-\tilde{q}\tilde{\alpha})\psi_{1j}
-(\tilde{\beta}\lambda_j+2\tilde{\beta}_x+\tilde{q}\tilde{\beta})\psi_{2j},
\end{array}\right.\end{equation}
where $1\leq j\leq N$, $\tilde{q}$, $\tilde{r}$, $\tilde{\alpha}$,
$\tilde{\beta}$ respectively denote $q$, $r$, $\alpha$, $\beta$
under the symmetry constraint (\ref{c10`}), and $\tilde{q}_x$,
$\tilde{r}_x$, $\tilde{\alpha}_x$, $\tilde{\beta}_x$ are given by
the following identities
$$\left\{\begin{array}{l}
\tilde{q}_x=<\Lambda\Psi_1, \Phi_2>-<\Psi_1, \Phi_2>^{2}+<\Psi_1,
\Phi_1>-<\Psi_2, \Phi_2>+\frac{1}{4}\psi_{N+1}(<\Psi_1,
\Phi_3>+<\Psi_3, \Phi_2>),\\
\tilde{r}_x=<\Psi_2, \Phi_1>-\frac{1}{2}<\Psi_1, \Phi_2>(<\Psi_1,
\Phi_1>-<\Psi_2, \Phi_2>)+\frac{1}{2}<\Psi_1,
\Phi_2>\phi_{N+1}\psi_{N+1},\\
\tilde{\alpha}_x=\frac{1}{4}(<\Psi_2, \Phi_3>-<\Psi_3,
\Phi_1>)+\frac{1}{2}<\Psi_1, \Phi_2>\phi_{N+1},\\
\tilde{\beta}_x=\frac{1}{4}(<\Psi_1, \Phi_3>+<\Psi_3,
\Phi_2>)-\frac{1}{8}<\Psi_1, \Phi_2>\psi_{N+1}.\end{array}\right.$$
Thus, the constrained system (\ref{t2-3}) becomes
\begin{equation}\left\{\begin{array}{l}
\Phi_{1, t_2}=\frac{1}{2}(-\Lambda^{2}+<\Lambda\Psi_1,
\Phi_2>+<\Psi_1, \Phi_1>-<\Psi_2, \Phi_2>)\Phi_1+\frac{1}{2}
[(<\Psi_1, \Phi_1>-<\Psi_2,
\Phi_2>)\Lambda\\\qquad\qquad-\phi_{N+1}\psi_{N+1}\Lambda+2<\Psi_2,
\Phi_1>]\Phi_2+\frac{1}{2}(2\phi_{N+1}\Lambda-<\Psi_2,
\Phi_3>+<\Psi_3, \Phi_1>)\Phi_3=\frac{\partial
H_2}{\partial\Psi_1},\\
\Phi_{2, t_2}=(\Lambda+<\Psi_1,
\Phi_2>)\Phi_1+\frac{1}{2}(\Lambda^{2}-<\Lambda\Psi_1,
\Phi_2>-<\Psi_1, \Phi_1>+<\Psi_2, \Phi_2>)\Phi_2+\frac{1}{4}
(\psi_{N+1}\Lambda\\\qquad\qquad+2<\Psi_1, \Phi_3>+2<\Psi_3,
\Phi_2>)\Phi_3=\frac{\partial H_2}{\partial\Psi_2},\\
\Phi_{3, t_2}=\frac{1}{4}(\psi_{N+1}\Lambda+2<\Psi_1,
\Phi_3>+2<\Psi_3,
\Phi_2>)\Phi_1-\frac{1}{2}(2\phi_{N+1}\Lambda-<\Psi_2,
\Phi_3>+<\Psi_3, \Phi_1>)\Phi_2\\\qquad\qquad=\frac{\partial
H_2}{\partial\Psi_3},\\
\phi_{N+1, t_2}=\frac{1}{2}\phi_{N+1}<\Lambda\Psi_1,
\Phi_2>+\frac{1}{4}(<\Lambda\Psi_2, \Phi_3>-<\Lambda\Psi_3,
\Phi_1>)=\frac{\partial H_2}{\partial\Psi_{N+1}},\\
\Psi_{1, t_2}=\frac{1}{2}(\Lambda^{2}-<\Lambda\Psi_1,
\Phi_2>-<\Psi_1, \Phi_1>+<\Psi_2, \Phi_2>)\Psi_1-(\Lambda+<\Psi_1,
\Phi_2>)\Psi_2+\frac{1}{4}(\psi_{N+1}\Lambda\\\qquad\qquad+2<\Psi_1,
\Phi_3>+2<\Psi_3, \Phi_2>)\Psi_3=-\frac{\partial
H_2}{\partial\Phi_1},\\
\Psi_{2, t_2}=-\frac{1}{2} [(<\Psi_1, \Phi_1>-<\Psi_2,
\Phi_2>)\Lambda-\phi_{N+1}\psi_{N+1}\Lambda+2<\Psi_2,
\Phi_1>]\Psi_1+\frac{1}{2}(-\Lambda^{2}+<\Lambda\Psi_1,
\Phi_2>\\\qquad\qquad+<\Psi_1, \Phi_1>-<\Psi_2,
\Phi_2>)\Psi_2-\frac{1}{2}(2\phi_{N+1}\Lambda-<\Psi_2,
\Phi_3>+<\Psi_3, \Phi_1>)\Psi_3=-\frac{\partial
H_2}{\partial\Phi_2},\\
\Psi_{3, t_2}=-\frac{1}{2}(2\phi_{N+1}\Lambda-<\Psi_2,
\Phi_3>+<\Psi_3,
\Phi_1>)\Psi_1-\frac{1}{4}(\psi_{N+1}\Lambda+2<\Psi_1,
\Phi_3>+2<\Psi_3, \Phi_2>)\Psi_2\\\qquad\qquad=\frac{\partial
H_2}{\partial\Phi_3},\\
\psi_{N+1, t_2}=<\Lambda\Psi_1, \Phi_3>+<\Lambda\Psi_3,
\Phi_2>-\frac{1}{2}\psi_{N+1}<\Lambda\Psi_1, \Phi_2>=\frac{\partial
H_2}{\partial\phi_{N+1}},\end{array}\right.\end{equation} where
Hamiltonian function is as follows
\begin{eqnarray*}
H_2&=&-\frac{1}{2}(<\Lambda^{2}\Psi_1, \Phi_1>-<\Lambda^{2}\Psi_2,
\Phi_2>)+\frac{1}{2}<\Lambda\Psi_1, \Phi_2>(<\Psi_1,
\Phi_1>-<\Psi_2, \Phi_2>)\\&& +<\Lambda\Psi_2,
\Phi_1>-\frac{1}{2}\phi_{N+1}\psi_{N+1}<\Lambda\Psi_1,
\Phi_2>+\frac{1}{4}\psi_{N+1}(<\Lambda\Psi_2,
\Phi_3>-<\Lambda\Psi_3, \Phi_1>)\\&& +<\Psi_2, \Phi_1><\Psi_1,
\Phi_2>-\frac{1}{2}(<\Psi_2, \Phi_3>-<\Psi_3, \Phi_1>)(<\Psi_1,
\Phi_3>+<\Psi_3, \Phi_2>)\\&&
+\phi_{N+1}(<\Lambda\Psi_1,
\Phi_3>+<\Lambda\Psi_3, \Phi_2>)+\frac{1}{4}(<\Psi_1,
\Phi_1>-<\Psi_2, \Phi_2>)^{2}.\end{eqnarray*}

Let's  construct integrals of motion for (\ref{c12}). An obvious
equality $(\tilde{V}^{2})_x=[\tilde{U}, \tilde{V}^{2}]$ leads to
\begin{equation}\label{cF}
F_x=(\frac{1}{2}Str\tilde{V}^{2})_x=
\frac{d}{dx}(\tilde{A}^{2}+\tilde{B}\tilde{C}+2\tilde{\rho}\tilde{\delta})=0,
\end{equation}
that is to say,  F is a generating function of integrals of motion
for the constrained spatial system (\ref{c12}). Since
$F=\sum\limits_{n\geq0}F_n\lambda^{-n}$, we obtain the following
expressions
$$F_n=\sum\limits_{i=0}^{n}(\tilde{A}_i\tilde{A}_{n-i}+\tilde{B}_i\tilde{C}_{n-i}
+2\tilde{\rho}_i\tilde{\delta}_{n-i}).$$ Using (\ref{c13}), we get
\begin{eqnarray}\label{c15}
F_0&=&\frac{1}{4}, \quad F_1=F_2=0,\nonumber\\
F_3&=&-\frac{1}{2}(<\Lambda\Psi_1, \Phi_1>-<\Lambda\Psi_2,
\Phi_2>)-\frac{1}{4}(<\Psi_2, \Phi_3>-<\Psi_3,
\Phi_1>)\psi_{N+1}\nonumber\\&&+\frac{1}{2}(<\Psi_1,
\Phi_1>-<\Psi_2, \Phi_2>-\phi_{N+1}\psi_{N+1})<\Psi_1,
\Phi_2>+<\Psi_2, \Phi_1>\nonumber\\&&+\phi_{N+1}(<\Psi_1,
\Phi_3>+<\Psi_3, \Phi_2>)=H_1,\nonumber\\
F_4&=&-\frac{1}{2}(<\Lambda^{2}\Psi_1, \Phi_1>-<\Lambda^{2}\Psi_2,
\Phi_2>)+\phi_{N+1}(<\Lambda\Psi_1, \Phi_3>+<\Lambda\Psi_3,
\Phi_2>)\nonumber\\&&+\frac{1}{2}(<\Psi_1, \Phi_1>-<\Psi_2,
\Phi_2>-\phi_{N+1}\psi_{N+1})<\Lambda\Psi_1, \Phi_2>+<\Lambda\Psi_2,
\Phi_1>\nonumber\\&&-\frac{1}{4}(<\Lambda\Psi_2,
\Phi_3>-<\Lambda\Psi_3, \Phi_1>)\psi_{N+1}+\frac{1}{4}(<\Psi_1,
\Phi_1>-<\Psi_2, \Phi_2>)^{2}\nonumber\\&&-\frac{1}{2}(<\Psi_2,
\Phi_3>-<\Psi_3, \Phi_1>)(<\Psi_1, \Phi_3>+<\Psi_3,
\Phi_2>)+<\Psi_2, \Phi_1><\Psi_1,
\Phi_2>,\nonumber\\
F_n&=&-\frac{1}{2}(<\Lambda^{n-2}\Psi_1,
\Phi_1>-<\Lambda^{n-2}\Psi_2,
\Phi_2>)+\phi_{N+1}(<\Lambda^{n-3}\Psi_1,
\Phi_3>+<\Lambda^{n-3}\Psi_3,
\Phi_2>)\nonumber\\&&+\frac{1}{2}(<\Psi_1, \Phi_1>-<\Psi_2,
\Phi_2>-\phi_{N+1}\psi_{N+1})<\Lambda^{n-3}\Psi_1,
\Phi_2>+<\Lambda^{n-3}\Psi_2, \Phi_1>\nonumber\\&&
-\frac{1}{4}(<\Lambda^{n-3}\Psi_2, \Phi_3>-<\Lambda^{n-3}\Psi_3,
\Phi_1>)\psi_{N+1}+\sum_{i=2}^{n-2}[\frac{1}{4}(<\Lambda^{i-2}\Psi_1,
\Phi_1>\nonumber\\&&-<\Lambda^{i-2}\Psi_2,
\Phi_2>)(<\Lambda^{n-i-2}\Psi_1, \Phi_1>-<\Lambda^{n-i-2}\Psi_2,
\Phi_2>)+<\Lambda^{i-2}\Psi_2,
\Phi_1>\nonumber\\&&<\Lambda^{n-i-2}\Psi_1,
\Phi_2>-\frac{1}{2}(<\Lambda^{i-2}\Psi_2,
\Phi_3>-<\Lambda^{i-2}\Psi_3, \Phi_1>)(<\Lambda^{n-i-2}\Psi_1,
\Phi_3>\nonumber\\&&+<\Lambda^{n-i-2}\Psi_3, \Phi_2>)],\quad
n\geq5.\end{eqnarray} Here $F_n(n\geq0)$ are all polynomials of
6N+2 dependent variables $\phi_{ij}$, $\psi_{ij}$, $\phi_{N+1}$ and
$\psi_{N+1}$, with $i=1, 2, 3$ and $j=1, \cdots, N$. Note that for
temporal part, $V_{t_n}=[V^{(n)}, V]$ is true. With the similar
discussion, we found that $F=\frac{1}{2}Str\tilde{V}^{2}$ is also a
generating function of integrals of motion for (\ref{ctime}).
Moreover, when  the symmetry constraint (\ref{c10`}) and new
independent variables (\ref{c23}) are considered, system
(\ref{ctime}) is constrained as follows
\begin{equation}\label{c18}\left\{\begin{array}{l}
\phi_{1j,
t_n}=(\sum\limits_{m=0}^{n}\tilde{A}_m\lambda_j^{n-m}+\frac{1}{2}\tilde{C}_{n+1})\phi_{1j}
+\sum\limits_{m=0}^{n}\tilde{B}_m\lambda_j^{n-m}\phi_{2j}+
\sum\limits_{m=0}^{n}\tilde{\rho}_m\lambda_j^{n-m}\phi_{3j},\quad 1\leq j\leq N,\\
\phi_{2j,
t_n}=\sum\limits_{m=0}^{n}\tilde{C}_m\lambda_j^{n-m}\phi_{1j}
-(\sum\limits_{m=0}^{n}\tilde{A}_m\lambda_j^{n-m}+\frac{1}{2}\tilde{C}_{n+1})\phi_{2j}
+\sum\limits_{m=0}^{n}\tilde{\delta}_m\lambda_j^{n-m}\phi_{3j},\quad 1\leq j\leq N,\\
\phi_{3j,
t_n}=\sum\limits_{m=0}^{n}\tilde{\delta}_m\lambda_j^{n-m}\phi_{1j}
-\sum\limits_{m=0}^{n}\tilde{\rho}_m\lambda_j^{n-m}\phi_{2j},\quad 1\leq j\leq N,\\
\phi_{N+1, t_n}=\frac{1}{2}\phi_{N+1}<\Lambda^{n-1}\Psi_1, \Phi_2>
+\frac{1}{4}(<\Lambda^{n-1}\Psi_2, \Phi_3>-<\Lambda^{n-1}\Psi_3,
\Phi_2>),\\
 \psi_{1j,
t_n}=-(\sum\limits_{m=0}^{n}\tilde{A}_m\lambda_j^{n-m}+\frac{1}{2}\tilde{C}_{n+1})\psi_{1j}
-\sum\limits_{m=0}^{n}\tilde{C}_m\lambda_j^{n-m}\psi_{2j}
+\sum\limits_{m=0}^{n}\tilde{\delta}_m\lambda_j^{n-m}\psi_{3j},\quad 1\leq j\leq N,\\
\psi_{2j,
t_n}=-\sum\limits_{m=0}^{n}\tilde{B}_m\lambda_j^{n-m}\psi_{1j}
+(\sum\limits_{m=0}^{n}\tilde{A}_m\lambda_j^{n-m}+\frac{1}{2}\tilde{C}_{n+1})\psi_{2j}
-\sum\limits_{m=0}^{n}\tilde{\rho}_m\lambda_j^{n-m}\psi_{3j},\quad 1\leq j\leq N,\\
\psi_{3j,
t_n}=-\sum\limits_{m=0}^{n}\tilde{\rho}_m\lambda_j^{n-m}\psi_{1j}
-\sum\limits_{m=0}^{n}\tilde{\delta}_m\lambda_j^{n-m}\psi_{2j},\quad
1\leq j\leq N,\\
\psi_{N+1, t_n}=<\Lambda^{n-1}\Psi_1, \Phi_3>+<\Lambda^{n-1}\Psi_3,
\Phi_2>-\frac{1}{2}\psi_{N+1}<\Lambda^{n-1}\Psi_1, \Phi_2>.
\end{array}\right.\end{equation}
After a direct calculation, we have
\begin{equation}\label{c19}\left\{\begin{array}{l}
\Phi_{1, t_n}=\frac{\partial F_{n+2}}{\partial\Psi_1},\quad \Phi_{2,
t_n}=\frac{\partial F_{n+2}}{\partial\Psi_2},\quad \Phi_{3,
t_n}=\frac{\partial F_{n+2}}{\partial\Psi_3},\quad \phi_{N+1,
t_n}=\frac{\partial F_{n+2}}{\partial\Psi_{N+1}},\\
\Psi_{1, t_n}=-\frac{\partial F_{n+2}}{\partial\Phi_1},\quad
\Psi_{2, t_n}=-\frac{\partial F_{n+2}}{\partial\Phi_2},\quad
\Psi_{3, t_n}=\frac{\partial F_{n+2}}{\partial\Phi_3},\quad
\psi_{N+1, t_n}=\frac{\partial
F_{n+2}}{\partial\Phi_{N+1}},\end{array}\right.\end{equation} which
shows that constrained system (\ref{c18}) is a super Hamiltonian
system.

In what follows, for 6N+2 dimensional super Hamiltonian systems
(\ref{c12}) and (\ref{c19}), we find 3N+1 integrals of
motion. It is natural to find that
\begin{equation}\label{c22}
f_k=\psi_{1k}\phi_{1k}+\psi_{2k}\phi_{2k}+\psi_{3k}\phi_{3k},\quad
1\leq k\leq N,\end{equation} are integrals of motion for constrained
systems (\ref{c12}) and (\ref{c19}). Therefore, for constrained
systems (\ref{c12}) and (\ref{c19}), we choose 3N+1 integrals of
motion
\begin{equation}\label{c3N+1}
f_1, \cdots, f_N, F_3, F_4, \cdots, F_{2N+3}.\end{equation} After a
simple calculation, we get
\begin{equation}\label{c21}
\{F_m, F_{n+2}\}=\frac{\partial}{\partial t_n}F_m=0,\end{equation}
where  Poisson bracket is defined by
\begin{equation}\label{c20}
\{F, G\}=\sum\limits_{i=1}^{3}\sum\limits_{j=1}^{N}(\frac{\partial
F}{\partial\phi_{ij}}\frac{\partial
G}{\partial\psi_{ij}}-(-1)^{p(\phi_{ij})p(\psi_{ij})}\frac{\partial
F}{\partial\psi_{ij}}\frac{\partial
G}{\partial\phi_{ij}})+\frac{\partial
F}{\partial\phi_{N+1}}\frac{\partial
G}{\partial\psi_{N+1}}+\frac{\partial
F}{\partial\psi_{N+1}}\frac{\partial
G}{\partial\phi_{N+1}}.\end{equation} The identity (\ref{c21}) means
that $\{F_m\}_{m\geq0}$ are in involution. The property of
involution among $\{f_k\}_{k=1}^{N}$ is obvious.
 About the independence of
$\{f_k\}_{k=1}^{N}$ and $\{F_m\}_{m=3}^{2N+3}$, we can refer to the
proof of Proposition 1 in \cite{HeYuZhouCheng}. Thus we obtain the
following theorem
\begin{Theorem}
The constrained systems (\ref{c12}) and (\ref{c19}) are Liouville
integrable super Hamiltonian systems, whose integrals of motion are
given by (\ref{c3N+1}).\end{Theorem}

\section{Conclusions and Discussions}

In this paper, the cKdV system is successfully extended to the super
one. For new system, its super Hamiltonian structure is expressed in
the form of (\ref{c7}). In our previous papers
\cite{HeYuZhouCheng}-\cite{YuHeMaCheng},the binary
nonlinearization has been applied to the super AKNS system and the super Dirac
system. For the super AKNS system, two kinds of nonlinearization of
Lax pairs, including nonlinearization under an explicit symmetry
constraint\cite{HeYuZhouCheng} and nonlinearization under an
implicit symmetry constraint\cite{YuHanHe}, have been considered
respectively. And for the super Dirac system, we only consider
binary nonlinearization under an explicit symmetry
constraint\cite{YuHeMaCheng}. From these three kinds of
nonlinearization of Lax pairs, the symmetry constraint
is either implicit or explicit. The novelty of the  constraint
(\ref{c10`}) for the super cKdV system is due to the combination of
the explicit constraint for even potentials $(q,r)$ and the implicit
constraint for odd potentials $(\alpha,\beta)$. Such combination
will make the process of binary nonlinearization complex. It is
highly non-trivial to solve $(\alpha,\beta)$ from implicit
constraints (\ref{c10`}) because it is related to a coupled
differential equations with variable coefficients. We introduce two
new odd variables (\ref{c23}) following the technique of implicit
constraint\cite{mali}. Thus, the spatial part and temporal parts of
the super cKdV system are nonlinearized respectively to the
constrained spatial system (\ref{c12}) and to the constrained
temporal system (\ref{c19}). Then, we see that systems (\ref{c12})
and (\ref{c19}) are super Hamiltonian systems. Furthermore,
constrained systems (\ref{c12}) and (\ref{c19}) are integrable in
the Liouville sense.

However, we are not able to do this for supersymmetric cKdV system.
Because spectral matrix of  supersymmetric cKdV system can not be
described by a certain Lie super algebra. In a word, how to make
nonlinearization of supersymmetric cKdV system is an interesting
problem. Furthermore, it is also an interesting problem to find an explicit solution
of the super finite dimensional integrable system. We shall consider these problems in
the future.

{\bf Acknowledgments}

 This work is supported by the Hangdian
Foundation KYS075608072 and KYS075608077, NSF of China under grant
number 10971109 and 11001069,
 and Program for NCET under Grant
  No.NCET-08-0515.  We thank anonymous referees for their valuable
suggestions and pertinent criticisms.


\begin{thebibliography}{99} 
\bibitem{Gurses1} M. G$\ddot{u}$rses, $\ddot{O}$. O$\check{g}$uz, A
super AKNS scheme, Phys. Lett. A 108 (1985), no. 9, 437-440.
\bibitem{Lizhang1} Y. S. Li, L. N.
Zhang, Super AKNS scheme and its infinite conserved currents, Nuovo.
Cimento A 93 (1986), no. 2, 175-183.
\bibitem{Lizhang2} Y. S. Li, L. N. Zhang, Hamiltonian structure of the super evolution equation, J. Math. Phys. 31
(1990), no. 2, 470-475.



\bibitem{Kupershmidt} B. A.
Kupershmidt, A super Korteseg-de Vries equation: an integrable
system, Phys. Lett. A 102 (1984), no. 5-6, 213-215.
\bibitem{Mathieu} P. Mathieu, Supersymmetric extension of the Korteweg-de Vries equation, J. Math. Phys. 29 (1988), no. 11, 2499-2506.
\bibitem{Ashok} D. Ashok, R. Shibaji, The zero curvature formulation of the sKdV equations, J. Math. Phys. 31 (1990), no. 9, 2145-2149.
\bibitem{Shaw1} J. C. Shaw, M. H. Tu, Binary Darboux-B$\ddot{a}$cklund transformations for the Manin-Radul super KdV hierarchy,
J. Math. Phys. 39 (1998), no. 9, 4773-4784.


\bibitem{Manin} Yu. I. Manin, A. O. Radul, A supersymmetric extension of the Kadomtsev-Petviashvili hierarchy, Comm. Math. Phys. 98
(1985) 65-77.
\bibitem{Yufeng} F. Yu, Bi-Hamiltonian structure of super KP hierarchy, J. Math. Phys. 33 (1992), no. 9, 3180-3189..
\bibitem{McArthur2} I. N. McArthur, Two-reduction of the super-KP hierarchy, Comm. Math. Phys. 159 (1994)
121-131.
\bibitem{Shaw2} J. C. Shaw, M. H. Tu, Hamiltonian structures of generated Manin-Radul super-KdV and constrained super KP hierarchies,
J. Math. Phys. 40 (1999), no. 6, 3021-3034.

\bibitem{Liu1} Q. P. Liu, Darboux transformation for the Manin-Radul supersymmetry KdV equation, M. Ma$\tilde{n}$as, Phys. Lett. B 394
(1997) 337-342.
\bibitem{Liu2} Q. P. Liu, Darboux transformations for super-symmetric KP hierarchies, Phys. Lett. B 485 (2000) 293-300.
\bibitem{Siddiq} M. Siddiq, M. Hassan, U. Saleem, On Darboux transformation of the supersymmetric sine-Gordon equation,
J. Phys. A: Math. Gen. 39 (2006) 7313-7318.


\bibitem{Popowicz} Z. Popowicz, Odd bihamiltonian structure of new supersymmetric N=2, 4 Korteweg de Vries equation and odd SUSY Virasoro-like
algebra, Phys. Lett. B 459 (1999) 150-158.
\bibitem{Manuel} F. R. Manuel, Dynamical symmetries, bi-Hamiltonian structures, and superintegrable n=2 systems,
J. Math. Phys. 41 (2000), no. 4, 2121-2134.
\bibitem{Kersten} P. H. M. Kersten, A. S. Sorin, Bi-Hamiltonian structure of the N-2 supersymmetric $\alpha$=1 KdV hierarchy,
Phys. Lett. A 300 (2002) 397-406.




\bibitem{Mathieu} S. Bourque, P. Mathieu, The Painlev$\acute{e}$ analysis for N=2 super Korteweg-de Vries equations, J. Math. Phys. 42
(2001), no.8, 3517-3539.

 \bibitem{HeYuZhouCheng} J. S. He, J. Yu, R. G. Zhou, Y. Cheng,
 Binary nonlinearization of the super AKNS system, Modern Phys. Lett. B 22 (2008), no. 4, 275-288.
 \bibitem{YuHanHe} J. Yu, J. W. Han, J. S. He, Binary nonlinearization of the super AKNS system under an implicit symmetry constraint,
J. Phys. A: Math. Theor. 42 (2009) 465201(10pp).
\bibitem{YuHeMaCheng} J. Yu, J. S. He, W. X. Ma, Y. Cheng, The Bargmann symmetry constraint and binary nonlinearization of the super Dirac systems,
Chin. Ann. Math. B 31 (2010), no. 3, 361-372.
\bibitem{Cao1} C. W. Cao, A cubic system which generates Bargmann potential and N-gap potential, Chin. Quart. J. Math. 3 (1988), no. 1, 90-96.
\bibitem{Ma1} W. X. Ma, W. Strampp, An explicit symmetry constraint for the Lax pairs and the adjoint Lax pairs of AKNS systems,
Phys. Lett. A 185 (1994), no. 3, 277-286.
\bibitem{Lizeng} Y. B. Zeng, Y. S. Li, The constraints of potentials and the finite-dimensional integrable systems, J. Math. Phys. 30
(1989), no. 8, 1679-1689.

\bibitem{Mazhou} W. X. Ma, R. G. Zhou, Adjoint symmetry constraints leading to binary nonlinearization, J. Nonlinear Math. Phys. 9
(2002) 106-126.
\bibitem{Xuxixiang} X. X. Xu, A generalized Wadati-Konno-Ichikawa hierarchy and its binary nonlinearization by symmetry constraints,
Chaos, Solitons and Fractals 15 (2003) 475-486.
\bibitem{Zhouruguang} R. G. Zhou, Nonlinearizations of spectral problems of the nonlinear Schr$\ddot{o}$dinger equation and the real-valued
modified Korteweg-de Vries equation, J. Math. Phys. 48 (2007), no.
1, 013510.
\bibitem{HirotaSatsuma} R. Hirota, J. Satsuma, Soliton solutions of
a coupled Korteweg-de Vries equation, Phys. Lett. A 85 (1981), no.
8-9, 407-408.

\bibitem{Cao1990} C. W. Cao, X. G. Geng, C. Neumann and Bargmann systems associated with the coupled KdV soliton hierarchy,
J. Phys. A 23 (1990), no. 18, 4117-4125.
\bibitem{Qin2004} Z. Y. Qin, Darboux transformations of the cKdV
hierarchy with sources, Chaos Solitons Fractals 21 (2004), no. 4,
989-998.
\bibitem{Hu} X. B. Hu, An approach to generate superextensions of integrable systems, J. Phys. A 30 (1997) no. 2, 619-632.
\bibitem{Maqin} W. X. Ma, J. S. He, Z. Y. Qin, A supertrace identity
and its applications to superintegrable systems, J. Math. Phys. 49
(2008), no. 3, 033511.
\bibitem{mali}Y. S. Li, W. X. Ma, Binary nonlinearization of AKNS spectral problem under higher-order symmetry
constraints, Chaos, Solitons and Fractals 11(2000), 697-710.


















\end{thebibliography}
\end{document}